\documentclass[]{article}
\usepackage{natbib}
\usepackage{amsmath}
\usepackage[dvips]{graphicx}
\usepackage{psfrag}
\usepackage{tikz}

\newcommand{\utau}{\tau^*}
\newcommand{\rme}{\mathrm{e}}

\newcommand{\Ut}{U_t}
\newcommand{\Ud}{U_\delta}
\newcommand{\vect}[1]{\mathbf{#1}}
\newcommand{\avg}[1]{\langle {#1} \rangle}
\newcommand{\Uw}{\mathcal{U}_w}
\newcommand{\T}{\mathcal{T}}
\newcommand{\PR}{\mathcal{P}_{req}}
\newcommand{\pipe}[1]{\left.#1\right|}
\newcommand{\p}{\partial}

\newcommand*\mycirc[1]{%
\begin{tikzpicture}
\node[draw,circle,inner sep=1pt] {#1};
\end{tikzpicture}}

\begin{document}

\title{The laminar generalized Stokes layer \\ and turbulent drag reduction}
\author{Maurizio Quadrio \\ \small Dipartimento di Ingegneria Aerospaziale del Politecnico di Milano \\ \small via La Masa 34, 20156 Milano, Italy \\ \\ Pierre Ricco \\ \small Department of Mechanical Engineering, King's College London, \\ \small Strand, London, WC2R 2LS United Kingdom
}
\date{}

\maketitle

\begin{abstract}

This paper considers plane channel flow modified by waves of spanwise velocity applied at the wall and travelling along the streamwise direction. Laminar and turbulent regimes for the streamwise flow are both studied. 

When the streamwise flow is laminar, it is unaffected by the spanwise flow induced by the waves. This flow is a thin, unsteady and streamwise-modulated boundary layer that can be expressed in terms of the Airy function of the first kind. We name it the generalized Stokes layer because it reduces to the classical oscillating Stokes layer in the limit of infinite wave speed. 

When the streamwise flow is turbulent, the laminar generalized Stokes layer solution describes well the space-averaged turbulent spanwise flow, provided that the phase speed of the waves is sufficiently different from the turbulent convection velocity, and that the time scale of the forcing is smaller than the life time of the near-wall turbulent structures. Under these conditions, the drag reduction is found to scale with the Stokes layer thickness, which renders the laminar solution instrumental for the analysis of the turbulent flow. 

A classification of the turbulent flow regimes induced by the waves is presented by comparing parameters related to the forcing conditions with the space and time scales of the turbulent flow.

\end{abstract}

\section{Introduction}
\label{sec:introduction}

The control of wall-bounded turbulent flows with the aim of reducing the wall-shear stress is an important and challenging topic in modern fluid mechanics. Such a reduction has enormous beneficial effects for engineering flow systems, such as less consumed fuel in aeronautical applications or for propelling gas and oil along pipelines. Feedback-control techniques \citep[see][for a recent review]{kasagi-suzuki-fukagata-2009} have recently seen their first experimental verification by \cite{yoshino-suzuki-kasagi-2008}, but still yield quite limited performance. Open-loop (predetermined) techniques, on the other hand, present much larger drag reduction rates, at the expense of a significant energy input. An example of open-loop strategy is the modification of wall turbulence by spanwise-traveling waves that produce large-scale spanwise forcing, either by a wall motion or a body force (see \cite{karniadakis-choi-2003} for a review). Such waves are effective in reducing the wall friction, as reported by numerical investigations \citep{du-karniadakis-2000,du-symeonidis-karniadakis-2002,zhao-wu-luo-2004} and by the experimental study by \cite{itoh-etal-2006}. 

In this paper, we focus on a different kind of traveling waves, i.e. streamwise-traveling waves of spanwise wall velocity. The significant effects of these waves on the friction drag has been recently reported by \cite{quadrio-ricco-viotti-2009} (hereinafter referred to as QRV09) through DNS of a plane channel flow. Their drag reduction properties have recently been confirmed by \cite{quadrio-etal-2009} through an experimental study of a turbulent pipe flow.

These waves are generated by the following space-time variation of the spanwise velocity $w_w^+$ at the wall:
\begin{equation}
w_w^+(x^+,t^+) = A^+ \cos \left( \kappa^+_x x^+ - \omega^+ t^+ \right),
\label{eq:space-time}
\end{equation}
where $x^+$ and $t^+$ denote the streamwise coordinate and time, $\kappa^+_x$ is the streamwise wavenumber, $\omega^+$ is the frequency, and $A^+$ is the amplitude. The $+$ sign indicates scaling by viscous units, i.e. by the friction velocity $u_\tau$ and the kinematic viscosity $\nu^*$. The traveling waves generalize the well-known spanwise oscillating-wall technique, for which $\kappa^+_x=0$, and the stationary streamwise-modulated spanwise oscillations, studied by \cite{viotti-quadrio-luchini-2009}, for which $\omega^+=0$. At a given $A^+$, the friction drag has been found by QRV09 to decrease for almost all the  $\kappa^+_x - \omega^+$ pairs. At $A^+=12$, full relaminarization is obtained at a friction-velocity Reynolds number $Re_\tau=100$. The maximum drag reduction is obtained by forward-traveling waves with a slow phase speed, $\Ut^+ \equiv \omega^+ / \kappa^+_x \approx 2$. 
The drag however increases by more than 20\% for waves with phase speeds comparable with the convection velocity $U_w^+$ of the near-wall coherent structures \citep{kim-hussain-1993}, i.e. for $\Ut^+ \approx 10$. Another interesting property is that large drag reductions are achieved with exceptionally low energetic expenditures: the energy required by the traveling waves can be several times smaller than the energy saved thanks to the reduced friction. 

Although the details are still unknown, the drag reduction effect is believed to be induced by the thin transversal boundary layer engendered by the wall waves, which are both unsteady and streamwise-modulated. This spanwise flow can be viewed as a generalized Stokes layer and it will be referred to as GSL in the following. 
In the special case of the oscillating wall, the spanwise turbulent flow averaged along the homogeneous (streamwise and spanwise) directions shows close agreement with the corresponding laminar solution \citep{choi-xu-sung-2002}, i.e. the unsteady Stokes layer produced by sinusoidal wall oscillations beneath a still fluid. For the standing-wave flow investigated by \cite{viotti-quadrio-luchini-2009}, the laminar solution also agrees well with the time-averaged spanwise turbulent profile. These laminar solutions have been useful for the prediction of relevant quantities related to the transversal shearing action in the turbulent regime, such as the spanwise velocity profile during the initial phase of the oscillation \citep{quadrio-ricco-2003} and the power spent for oscillating the wall against the frictional resistance of the fluid \citep{ricco-quadrio-2008}.

The existence of analytical solutions for the transversal boundary layers in the special oscillating-wall and steady-waves cases and the usefulness of such solutions for the prediction and understanding of drag reduction lead us to the central questions addressed in this paper. The analysis of the standing-wave laminar flow is first extended to include the unsteady effects induced by the wall forcing (\ref{eq:space-time}). Then, we ask ourselves whether the spatio-temporal GSL solution can be helpful to study turbulent drag reduction by traveling waves. We find an analytical expression for the spatio-temporal GSL flow superimposed to a laminar Poiseuille streamwise flow. This solution reduces to the classical Stokes solution when $\kappa_x^+ \rightarrow 0$. It is then shown that this laminar solution agrees well with the space-averaged spanwise turbulent flow under well-defined forcing conditions. The solution is further employed to compute the GSL thickness and the power spent to drive the wall waves. The study reveals that, under the conditions of agreement between the laminar solution and the spanwise turbulent flow, the drag reduction scales with the GSL thickness and that a minimum value of the thickness is required for the waves to yield drag reduction. 
The role of the thickness of the spanwise boundary layer is further highlighted by studying turbulent statistics. To this end, the database by QRV09 is expanded with several additional simulations where the streamwise pressure gradient is held constant, so that the modified flow is characterized by a well-defined value of $Re_\tau$, while the flow rate is left free to adapt to the new state. This approach allows expressing the flow quantities through the proper inner scaling. We also interpret physically the role of the traveling waves by comparing a combination of their wavelength and frequency to a typical time scale of the near-wall turbulence and their phase speed with the turbulent convection velocity.

The structure of the paper is as follows. For the laminar flow, the mathematical problem is formulated in \S\ref{sec:laminar-setup}. The spanwise momentum equation is cast in nondimensional form and simplified in \S\ref{sec:laminar-scaling}; different flow regimes are distinguished in \S\ref{sec:laminar-regimes}. The key assumption of thin GSL thickness is outlined in \S\ref{sec:laminar-thinlayer} and an analytical solution is found in terms of the Airy function of the first kind in \S\ref{sec:laminar-solution}; lastly, some quantities characterizing the GSL are introduced in \S\ref{sec:derived-quantities}.
For the turbulent flow, the differences between constant mass flux and constant streamwise pressure gradient simulations are discussed in \S\ref{sec:turbulent-scaling}. The GSL laminar solution is compared with the mean spanwise turbulent profile in \S\ref{sec:turbulent-comparison} and in \S\ref{sec:turbulent-correlation} the turbulent drag changes are correlated with the quantities computed through the GSL solution. Lastly, four classes of flow regimes are distinguished in \S\ref{sec:turbulent-regimes}. The final section \S\ref{sec:summary} contains a summary of the results. 

\section{Laminar flow}
\label{sec:laminar}

\begin{figure}
\centering
\psfrag{F}{Streamwise flow}
\psfrag{x}{$x^*$}
\psfrag{y}{$y^*$}
\psfrag{z}{$z^*$}
\psfrag{Ly}{$2 h^*$}
\psfrag{W}{$w_w^*=A^* \cos \left(\kappa_x^* x^* - \omega^* t^* \right)$}
\psfrag{U}{$\Ut^*$}
\psfrag{l}{$\lambda_x^*$}
\psfrag{d}{$\delta^*$}
\centering
\includegraphics[width=\textwidth]{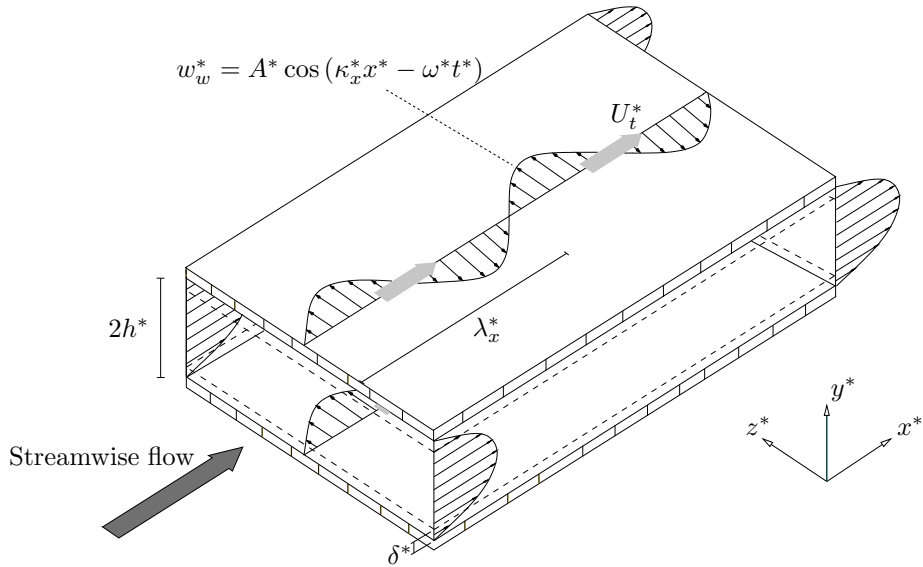}
\caption{Schematic of the physical domain for laminar channel flow with forward-traveling wall waves. The channel width is $2h^*$, $\delta^*$ is the GSL thickness, $\lambda_x^*= 2 \pi / \kappa_x^*$ is the streamwise wavelength of the wall forcing, and $\Ut^*$ is the phase speed.}
\label{fig:traveling wave}
\end{figure}

The flow induced in a laminar Poiseuille channel flow by the wall motion given by (\ref{eq:space-time}) is first studied. We consider a laminar incompressible flow driven by a constant pressure gradient between two indefinite parallel planes, separated by a distance $2 h^*$. The Cartesian coordinates $x^*$, $y^*$ and $z^*$ indicate the streamwise, wall-normal and spanwise directions, respectively, $t^*$ denotes time, and the symbol $^*$ indicates dimensional quantities. The flow is governed by the incompressible Navier--Stokes equations
\begin{equation}
\label{eq:ns-continuity}
\nabla \cdot \vect{V}^* = 0
\end{equation}
\begin{equation}
\label{eq:ns-momentum}
\frac{\p \vect{V}^*}{\p t^* } + \left( \vect{V^*} \cdot \nabla \right) \vect{V^*} = -\frac{1}{\rho^*} \nabla p^* + \nu^* \nabla^2 \vect{V^*},
\end{equation}
where $\vect{V^*}=\{u^*,v^*,w^*\}$ is the velocity vector with components along $x^*$, $y^*$ and $z^*$, $p^*$ is the pressure, $\rho^*$ and $\nu^*$ are the density and the kinematic viscosity of the fluid, and $\nabla=\{\p/\p x^*,\p/\p y^*,\p/\p z^*\}$. At the walls, i.e. at $y^*$=0 and $y^*=2h^*$, the spanwise velocity component takes the form of a traveling wave, so that the following boundary conditions are imposed
\begin{equation}
\label{eq:bc}
u^*=v^*=0, \qquad w^* = A^* \Re \left[ \rme^{i \kappa^*_x \left( x^* - \Ut^* t^* \right) } \right],
\end{equation}
where $\Re$ indicates the real part, $\kappa^*_x$ is the streamwise wavenumber, and $\Ut^*=\omega^*/\kappa_x^*$ is the phase speed of the traveling wave, where $\omega^*$ is the frequency. $\Ut^*$ can be positive (forward-traveling wave), null (standing wave) or negative (backward-traveling wave).
The flow presents a symmetry with respect to the origin of the $\omega^* - \kappa_x^*$ plane since the forcing (\ref{eq:bc}) is invariant to a change of $(\omega^*,\kappa_x^*)$ into $(-\omega^*,-\kappa_x^*)$. A schematic of the physical domain for the case of forward-traveling waves is shown in figure \ref{fig:traveling wave}.

%%%%%%%%%%%%%%%%%%%%%%%%%%%%%%%%%%%%%%%%%%%%%%%%%%%%%%%%%%%%%%%
\subsection{The laminar solution of the generalized Stokes problem}
\label{sec:laminar-setup}

The system (\ref{eq:ns-continuity})--(\ref{eq:ns-momentum}) can be simplified as follows. All terms involving the $z^*$ derivatives are null because the non-homogeneous boundary condition (\ref{eq:bc}) depends on the sole coordinate $x^*$ and there is no pressure gradient along $z^*$. Analogously to the classical channel flow, from this simplification and the use of (\ref{eq:ns-continuity}) it follows that the $x$- and $y$-momentum equations become independent of $w^*$. The streamwise flow is thus described by the steady parabolic velocity profile of the plane Poiseuille flow, $v^* = 0$ everywhere in the channel, and $w^*$ satisfies the $z$-momentum equation
\begin{equation}
\label{eq:w}
\frac{\p w^*}{\p t^*} + u^* \frac{\p w^*}{\p x^*} = \nu^* \left( \frac{\p^2 w^*}{\p x^{*2}} + \frac{\p^2 w^*}{\p y^{*2}} \right).
\end{equation}
This equation bears some resemblance with the one describing the laminar flow induced by an indefinite flat plate oscillating sinusoidally in time below a still fluid, a problem often referred to as the second Stokes problem \citep{batchelor-1967}, where the boundary condition is $w_w^* = A^* \Re \left( \rme^{i \omega^* t^* } \right)$. In this case, an unsteady boundary layer develops (usually referred to as the Stokes layer, TSL) and its thickness is inversely proportional to the square root of the frequency. Two terms of (\ref{eq:w}) are however absent in the equation for the classical Stokes problem: the second convective term on the left (which describes the one-way coupling between the Poiseuille parabolic profile $u^*(y^*)$ and the spanwise flow), and the diffusion term along $x^*$ (which is not null because the wall boundary condition for $w^*$ depends on $x^*$). The steady version of (\ref{eq:w}) was investigated by \cite{viotti-quadrio-luchini-2009} for a wall forcing by stationary waves, i.e. $w_w^* = A^* \Re \left( \rme^{i \kappa_x^* x^* } \right)$. A thin steady spatially-modulated viscous layer is generated (SSL) and its thickness is proportional to the cubic root of the streamwise wavelength. As expected in view of the analogy with the temporal and spatial Stokes problems, a thin viscous transversal boundary layer, the generalized Stokes layer, develops in the present traveling-wave case. The GSL is both unsteady and spatially modulated along the streamwise direction. We work under the hypothesis that the GSL thickness is much smaller than the distance between the channel walls (see discussion in \S\ref{sec:laminar-thinlayer}). 

%%%%%%%%%%%%%%%%%%%%%%%%%%%%%%%%%%%%%%%%%%%%%%%%%%%%%%%%%%%%%%%%%%%%%%%%
\subsection{Scaling and simplification of the spanwise momentum equation}
\label{sec:laminar-scaling}

Equation (\ref{eq:w}) is now nondimensionalized and expressed in a simplified form. The problem involves two distinct length scales. The first one is $\lambda_x^* = 2 \pi / \kappa_x^*$, the streamwise wavelength of the traveling wave. The second length scale is $\delta^*$, a measure of the GSL thickness. The scaled streamwise coordinate is therefore $x=x^*/\lambda_x^*=\mathcal{O}(1)$ and the scaled wall-normal coordinate is $y=y^*/\delta^*=\mathcal{O}(1)$.

Analogously, two velocity scales exist in the boundary layer. The first one is related to the streamwise flow within the layer: it can be taken as the maximum streamwise velocity across the layer, i.e. the Poiseuille flow velocity at the edge of the GSL. On defining $\Ud^* \equiv u^*(\delta^*)$, $u=u^*/\Ud^*=\mathcal{O}(1)$. 

By assuming that
\begin{equation}
\label{eq:thin-layer-assumption}
\delta^* \ll h^*,
\end{equation}
$u^*(y^*)$ can be legitimately expressed through a Taylor expansion for small $y^*$: $u^*(y^*) = u^*(0) + y^* \utau + \mathcal{O}(y^{*2}), \utau \equiv \pipe{\mbox{d} u^*/\mbox{d}y^*}_{y^*=0}$. The velocity scale becomes
\begin{equation}
\label{eq:udelta}
\Ud^* = \delta^* \utau,
\end{equation}
and $u=y$. The Poiseuille velocity profiles becomes equivalent to the Couette laminar profile. Henceforth, $u$ can therefore be thought of as a Couette laminar flow bounded at $y=0$ and unbounded as $y \rightarrow \infty$. Note that (\ref{eq:thin-layer-assumption}) will be later expressed in terms of $\nu^*$, $h^*$, $\lambda_x^*$ and the bulk streamwise velocity $U_b^*$ when an expression for $\delta^*$ is found (see \S\ref{sec:laminar-regimes}).
The spanwise velocity component scales with $A^*$, i.e. $w = w^*/A^* = \mathcal{O}(1)$, and the time is scaled by the period of the wall motion, i.e. $t=t^*\Ut^*/\lambda^*_x=\mathcal{O}(1)$. Upon substituting the scaled variables into (\ref{eq:w}), one finds
\begin{equation}
\label{eq:w-adim}
\frac{\p w}{\p t} + \frac{\Ud^*}{\Ut^*} y \frac{\p w}{\p x} = \frac{\nu^*}{\Ut^* \lambda_x^*} \frac{\p^2 w}{\p x^2} + \frac{\lambda_x^* \nu^*}{\Ut^* \left(\delta^*\right)^2} \frac{\p^2 w}{\p y^2}.
\end{equation}
The boundary conditions are $w = \Re \left[ \rme^{2 \pi i \left( x - t \right)} \right]$ at $y=0$, and $w = 0$ as $y \rightarrow \infty$. The variable $\xi=x-t$ may be introduced, so that $\p / \p x = \p / \p \xi$ and $\p/\p t = - \p/\p \xi$. Equation (\ref{eq:w-adim}) becomes
\begin{equation}
\label{eq:z-mom}
\left( y - \frac{\Ut^*}{\Ud^*} \right) \frac{\p w }{\p \xi} =
\frac{\nu^*}{\Ud^* \lambda_x^*} \frac{\p^2 w}{\p \xi^2} +
\frac{\lambda_x^* \nu^*}{\Ud^* \left(\delta^*\right)^2} \frac{\p^2 w}{\p y^2},
\end{equation}
and $w=\Re(\rme^{2 \pi i \xi})$ at $y=0$. 

%%%%%%%%%%%%%%%%%%%%%%%%%%%%%%%%%%%%%%%%%%%%%%%%%%%%%%%%%%%%%%%%%%%%%%%

\subsection{The three GSL flow regimes and the GSL thickness}
\label{sec:laminar-regimes}

Three flow regimes can be identified by considering the magnitude of the inertial terms in (\ref{eq:z-mom}) with respect to the $y$-diffusion term. We only discuss cases with $\lambda_x^* > 0$ as the flow presents a symmetry with respect to the origin of the $\omega^* - \kappa^*$ plane (see discussion at beginning of \S\ref{sec:laminar}).
In the limit
\begin{equation}
\label{eq:limit-temporal}
\delta^* \ll \Ut^*/\utau,
\end{equation}
the balance between inertial and viscous effects gives 
\begin{equation}
\label{eq:delta-stokes}
\delta^* = \mathcal{O}\left[ \left( \frac{\lambda_x^* \nu^*}{\Ut^*} \right)^{1/2} \right]
\end{equation}
(In this limit, the absolute value of the phase speed must be considered to obtain a real value for $\delta^*$.) By inserting (\ref{eq:delta-stokes}) into (\ref{eq:limit-temporal}), a condition on the phase speed for the oscillating-wall regime is obtained: $\Ut^* \gg ( \lambda_x^* \nu^* )^{1/3} {\utau}^{2/3}$. The phase speed is so high that the convection is due solely to the unsteadiness. The flow induced by spatially uniform temporal wall oscillations, i.e. the classical TSL, is recovered. In the opposite limit
\begin{equation}
\label{eq:limit-standing}
\delta^* \gg \Ut^*/\utau,
\end{equation}
the balance between inertial and viscous effects leads to
\begin{equation}
\label{eq:delta-standing}
\delta^* = \mathcal{O}\left[\left( \frac{\lambda_x^* \nu^*}{\utau}\right)^{1/3}\right].
\end{equation}
By substituting (\ref{eq:delta-standing}) into (\ref{eq:limit-standing}), the condition for the standing-wave regime is found: $\Ut^* \ll \left( \lambda_x^* \nu^* \right)^{1/3} \left( \utau \right)^{2/3}$. The phase speed is so low that the convection is due solely to the streamwise modulation. The flow induced by stationary waves, i.e. the SSL studied by \cite{viotti-quadrio-luchini-2009}, is obtained. 
The intermediate case occurs when the two inertia terms on the left side of (\ref{eq:z-mom}) are comparable; the unsteadiness and the streamwise modulation both contribute to the convection. The boundary-layer balance is
\begin{equation}
\label{eq:limit-traveling}
\frac{\left(\delta^*\right)^3 \utau}{\lambda_x^* \nu^*} =  \mathcal{O}\left[\frac{\left(\delta^*\right)^2 \Ut^*}{\lambda_x^* \nu^*} \right] =  \mathcal{O}(1),
\end{equation}
or $\delta^* = \mathcal{O}(\Ut^*/\utau)$. A characteristic phase speed
\begin{equation}
\label{eq:characteristic-speed}
U_{t,c}^* \equiv \left( \lambda_x^* \nu^* \right)^{1/3} \left( \utau \right)^{2/3}
\end{equation}
denotes the traveling-wave regime. The speed $U_{t,c}^*$ has the following physical interpretation. By substituting either (\ref{eq:delta-stokes}) or (\ref{eq:delta-standing}) into (\ref{eq:udelta}), one finds that $U_{t,c}^*=\mathcal{O}(\Ud^*)$. This means that the characteristic phase speed is comparable with the streamwise velocity at the outer edge of GSL when the traveling-wave regime occurs.

In the oscillating-wall regime, $\delta^*$ is given by (\ref{eq:delta-stokes}), or, equivalently, by $\delta^* = \mathcal{O}[(\nu^*/\omega^*)^{1/2}]$, which is the classical Stokes layer result. For the standing-wave regime, we recognize in (\ref{eq:delta-standing}) the $\lambda_x^{* 1/3}$-dependence of the SSL thickness, which is due to the coupling with the streamwise flow. A $1/3$-algebraic dependence of $\delta^*$ on the streamwise length scale (here the wavelength $\lambda_x^*$) is a recurrent fact whenever the inertial term is dictated by a uniform spanwise vorticity (here the laminar Couette flow). Classic examples include the trailing-edge laminar wake \citep{goldstein-1930}, and the flat-plate laminar boundary layer beneath an inviscid flow with high shear \citep{ting-1960}. In the traveling-wave regime, simplification of (\ref{eq:limit-traveling}) shows that both expressions for $\delta^*$, i.e. (\ref{eq:delta-stokes}) for the oscillating-wall regime and (\ref{eq:delta-standing}) for the standing-wave regime, are valid.

It is interesting to point out that the order of magnitude of $\delta^*$ may be extracted in another way. The balance between inertial and $y$-diffusion terms in (\ref{eq:z-mom}) may be rewritten as follows
\begin{equation}
\label{eq:cubic}
\left(\delta^*\right)^3  -  \frac{\Ut^*}{c_2 \utau} \left(\delta^*\right)^2 = \frac{c_1 \lambda_x^* \nu^*}{c_2 \utau},
\end{equation}
where $c_1=(\p^2 w/\p y^2)/(\p w /\p \xi) = \mathcal{O}(1)$, $c_2=y=\mathcal{O}(1)$.
In the classical Stokes problem, only $c_1$ is relevant ($c_2$=0), so that (\ref{eq:cubic}) leads to $\delta^* = \mathcal{O}[(\nu^*/\omega^*)^{1/2}]$. The order of magnitude for $\delta^*$ in the standing-wave regime, given by (\ref{eq:delta-standing}), is recovered by setting $c_2 \rightarrow \infty$, with $c_1/c_2 = \mathcal{O}(1)$. Although the solution of the cubic algebraic equation (\ref{eq:cubic}) \citep{dunham-1990} proves unsuitable to yield a simple expression for $\delta^*$ because of the nonlinear dependence on two order-one constants, it is instructive to inspect the properties of the discriminant $\Delta$ of (\ref{eq:cubic}), $\Delta = (c_1 \lambda_x^* \nu^*) \left( \Ut^*/(3 c_2 \utau) \right)^3/(c_2 \utau) + \left( c_1 \lambda_x^* \nu^*/(2 c_2 \utau) \right)^2$. When $\Delta=0$, $\Ut^*=-3 (c_1 \lambda_x^* \nu^*)^{1/3} (c_2 \utau)^{2/3}$, which is of the same order of $U_{t,c}^*$ in (\ref{eq:characteristic-speed}). Three real solutions of (\ref{eq:cubic}) exist (two of which are equal), and the only positive one is $\delta^*=(c_1 \lambda_x^* \nu^*/(4 c_2 \utau))^{1/3}$, which is the traveling-wave result found in (\ref{eq:delta-standing}). Setting $\Delta \gg 0$ implies that $\Ut^* \ll U_{t,c}^*$, which identifies the standing-wave regime. This is further confirmed by the only real solution of (\ref{eq:cubic}) being of the form (\ref{eq:delta-standing}). For $\Delta \ll 0$, the oscillating-wall regime is recovered, $\Ut^* \gg U_{t,c}^*$. Three real solutions exist, amongst which the only positive one is (\ref{eq:delta-stokes}). Therefore, $\Delta$ is not only a discriminant in mathematical terms, but it also allows distinguishing amongst the physical regimes.

%%%%%%%%%%%%%%%%%%%%%%%%%%%%%%%%%%%%%%
\subsection{The thin-layer assumption}
\label{sec:laminar-thinlayer}

Assumption (\ref{eq:thin-layer-assumption}) can be recast in a more convenient form in terms of the known quantities $\lambda_x^*$, $h^*$, $\nu^*$, $U_b^*$ by use of the expressions for $\delta^*$ found in \S\ref{sec:laminar-regimes}. Use of (\ref{eq:udelta}), (\ref{eq:delta-stokes}), (\ref{eq:delta-standing}), and the relation between the wall-shear stress and $Re_b= 2 U_b^* h^*/\nu^*$ for a laminar channel flow, i.e. $\nu^* \utau = 12 (U_b^*)^2/Re_b$, shows that (\ref{eq:thin-layer-assumption}) becomes
\begin{equation}
\label{eq:thin-layer-assumption-detailed}
\left.
\begin{array}{ll}
\mbox{Standing-wave regime}:      \lambda_x^*/h^*  \ll Re_b \\
\mbox{Oscillating-wall regime}:   \lambda_x^*/h^*  \ll \Ut^* h^* /\nu^* \ \mbox{or}  \ \omega^* \gg \nu^*/(h^*)^2 \\
\mbox{Traveling-wave regime}:     \lambda_x^*/h^*  \ll Re_b \ \mbox{or} \ \lambda_x^*/h^* \ll \Ut^* h^* /\nu^* 
\ \ \mbox{or} \ \ \omega^* \gg \nu^*/(h^*)^2.
\end{array}
\right\}
\end{equation}
As a summary, for the standing-wave and the traveling-wave regimes, the thin-layer approximation can be written as
\begin{equation}
\label{eq:assumptions-laminar-lambda}
\frac{\lambda_x^*}{h^*} \ll Re_b,
\end{equation}
while, for the oscillating-wall regime, it may be expressed as
\begin{equation}
\label{eq:assumptions-laminar-omega}
\frac{\omega^* (U_b^*)^2}{h^*} \gg \frac{1}{Re_b}.
\end{equation}
For a streamwise turbulent flow, studied in \S\ref{sec:turbulent}, the wall friction and $Re_b$ are related by $ \nu^* \utau/(U_b^*)^2 \approx 0.1228 Re_b^\alpha$, $\alpha=-0.25$ \citep{dean-1978}. This relationship is used in the present study because it has been verified to match very well DNS data up to $Re_\tau=2000$ \citep{marusic-joseph-mahesh-2007}. Equation (\ref{eq:assumptions-laminar-lambda}) therefore becomes
\begin{equation}
\label{eq:assumptions-turbulent}
\frac{\lambda_x^*}{h^*} \ll Re_b^{\alpha+2}.
\end{equation}
In QRV09's DNS of a turbulent channel flow and in the present simulations, $Re_b$ is always higher than 3000, so that (\ref{eq:assumptions-turbulent}) translates to $\lambda_x^*/h^* \ll 10^6$, which is amply verified because $\lambda_x^* / h^*$ varies between $\approx 1$ and $\approx 30$. In viscous units, (\ref{eq:assumptions-turbulent}) becomes $\lambda_x^+ \ll (U_b^+)^{\alpha+2} Re_\tau^{\alpha+3}$. The inequality (\ref{eq:assumptions-laminar-omega}) holds when the streamwise flow is turbulent because $\delta^*$ does not depend on $\tau^*$ in the oscillating-wall regime.

%%%%%%%%%%%%%%%%%%%%%%%%%%%%%%%%%%%%%%%%%%%%%%%%%%%%%%%%%%%%%%%%%%%%%%%%%%%
\subsection{Analytical expression for the GSL}
\label{sec:laminar-solution}

We now express the spanwise momentum equation (\ref{eq:z-mom}) in a more compact form to arrive at an analytical formula in terms of Airy function of the first kind (see \cite{abramowitz-stegun-1964} at page 446). The parameter $\overline{U}=\Ut^*/U_{t,c}^*$ may now be introduced. The standing-wave regime dominates when $\overline{U} \ll 1$, the oscillating-wall regime is effective when $\overline{U} \gg 1$, and $\overline{U} = \mathcal{O}(1)$ when the traveling-wave regime occurs. By using (\ref{eq:udelta}), equation (\ref{eq:z-mom}) becomes
\[
\left( y - \overline{U} \right) \frac{\p w }{\p \xi} =
\frac{\nu^*}{\delta^* \utau \lambda_x^*} \frac{\p^2 w}{\p \xi^2} +
\frac{\lambda_x^* \nu^*}{\utau \left(\delta^*\right)^3} \frac{\p^2 w}{\p y^2},
\]
which further simplifies to
\begin{equation}
\label{eq:z-mom-simplified-2}
i \Upsilon^{-3} (y - \tilde{U}) F(y) = \frac{\mbox{d}^2 F}{\mbox{d} y^2},
\end{equation}
on defining $\Upsilon \equiv (\delta^*)^{-1}(\lambda_x^* \nu^*/(2 \pi \utau) )^{1/3}$, $\tilde{U}=\overline{U}+2 \pi i \nu^*/(\lambda_x^* \utau \delta^*)$ and $w(\xi,y) = \Re \left[ F(y) \rme^{2 \pi i \xi} \right]$. Equation (\ref{eq:z-mom-simplified-2}) is subject to the boundary conditions
\begin{equation}
\label{eq:bc-z-mom-simplified-2}
F(0) = 1, \qquad \lim_{y \rightarrow \infty} F(y) = 0.
\end{equation}
The transformations $\tilde y = i ( \tilde{U} - y )/\Upsilon,  \tilde F(\tilde y) = F(i \Upsilon \tilde y + \tilde{U})$
lead to the Airy equation $\tilde y \tilde F(\tilde y) = \mbox{d}^2 \tilde F/\mbox{d} \tilde y^2$, whose solution can be expressed in terms of Airy function of the first kind, $\tilde F(\tilde y) = \theta \mbox{Ai}(\tilde y) + \gamma \mbox{Ai} \left(\tilde y \rme^{2 \pi i/3}\right)$, where $\theta$ and $\gamma$ are constants. By applying (\ref{eq:bc-z-mom-simplified-2}), it follows that $\theta=0$. The spanwise velocity profile becomes
\begin{equation}
\label{eq:airy-solution}
w^*(x^*,y^*,t^*) = A^*\Re \left\{ C \rme^{i (\kappa_x^* x^* - \omega^* t^*)}
\mbox{Ai}\left[ \rme^{\pi i/6} \left( \frac{\kappa_x^* \utau}{\nu^*} \right)^{1/3}
\left( y^* - \frac{\omega^*}{\kappa_x^* \utau} - \frac{i \kappa_x^* \nu^*}{\utau} \right) \right] \right\},
\end{equation}
where $C=\left\{\mbox{Ai}\left[ i \rme^{ i \pi /3} \left(\kappa_x^* \utau/\nu^*\right)^{1/3} (\omega^*/\kappa^* + i \kappa_x^* \nu^*)/\utau \right]\right\}^{-1}$ is a constant. The formula above simplifies to the one found by \cite{viotti-quadrio-luchini-2009} for the SSL when $\omega^*=0$ and $i \kappa_x^* \nu^*/\utau=0$. This latter condition follows from neglecting the streamwise diffusion, i.e. in the limit of large streamwise wavelength. Through an analysis analogous to the one for the thin-layer approximation, it can be shown that viscous effects along $x^*$ can be formally neglected when $\lambda_x^*/h^* \gg Re_b^{-1/2}$ for a laminar flow and when $\lambda_x^*/h^* \gg Re_b^{-(\alpha+1/2)}$ for a turbulent flow. 

Equation (\ref{eq:airy-solution}) is not defined when $\kappa_x^*=0$ (oscillating-wall flow); in this case, however, the analytic formula for the classical Stokes problem is valid:
\begin{equation}
\label{eq:stokes-flow}
w^*(x^*,y^*,t^*) = A^* \exp \left(-y^* \sqrt{\frac{\omega^*}{2 \nu^*}}\right) \cos \left(\omega^* t^* -y^* \sqrt{\frac{\omega^*}{2 \nu^*}} \right).
\end{equation}
It is therefore useful to verify that formula (\ref{eq:airy-solution}) for the traveling waves matches asymptotically the Stokes layer solution (\ref{eq:stokes-flow}) as $\kappa_x^* \rightarrow 0$. We study the cases with $\omega^*/\kappa_x^*>$0 (the analysis is analogous for negative ratios). In this limit, the argument $\zeta^*$ of the Airy function in (\ref{eq:airy-solution}) is unbounded
\begin{equation}
\label{eq:airy-argument}
\zeta^* \sim \rme^{-5 \pi i/6}\left( \frac{\kappa_x^* \utau}{\nu^*} \right)^{1/3}
\left(  \frac{\omega^*}{\kappa_x^* \utau} - y^* \right),
| \zeta^* | \rightarrow \infty.
\end{equation}
The following asymptotic formula therefore applies:
\begin{equation}
\label{eq:airy-asymptotic}
\mbox{Ai}(\zeta^*) \sim \frac{1}{2 \sqrt{\pi}} {(\zeta^*)}^{-1/4} \rme^{-2 {(\zeta^*)}^{3/2}/3} 
\sum_{k=0}^{\infty} (-1)^k c_k \left(\frac{2 {(\zeta^*)}^{-3/2}}{3}\right)^{-k}, 
| \zeta^* | \rightarrow \infty, |\mbox{arg}(\zeta^*)| < \pi,
\end{equation}
where $c_k$ are given in 10.4.58 at page 458 in \cite{abramowitz-stegun-1964}. By substituting (\ref{eq:airy-argument}) into (\ref{eq:airy-asymptotic}) and then into (\ref{eq:airy-solution}), one finds
\begin{equation}
\label{eq:airy-solution-asymptotic}
w^*=A^*\Re \left\{ 
\frac{C \rme^{-i \omega^* t^*} {(\kappa_x^*)}^{1/6} {(\nu^*)}^{1/12} \rme^{5 \pi i/24}}{2 
\sqrt{\pi} {(\omega^*)}^{1/4} {(\tau^*)}^{1/3}} \exp\left[ 
\frac{2 \rme^{-5 \pi i/4} i {(\omega^*)}^{3/2}}{3 \kappa_x^* \utau \sqrt{\nu^*}}  
\left(1 - \frac{\kappa_x^* y^* \utau}{\omega^*} \right)^{3/2}
\right]
\right\}.
\end{equation}
The last algebraic term in (\ref{eq:airy-solution-asymptotic}), once expanded by Taylor series with respect to $\kappa_x^*$, $\left(1 - \kappa_x^* y^* \utau/\omega^* \right)^{3/2} = 1 - 3 \kappa_x^* y^* \utau/(2 \omega^*) + ...$, can be substituted into (\ref{eq:airy-solution-asymptotic}) to obtain (\ref{eq:stokes-flow}).

We close this section by noting that equation (\ref{eq:w-adim}) also arises in the study of the stability of Couette flow as a simplification of the Orr-Sommerfeld equation \citep{orr-1907,marcus-1977}. The unknown variable $S=\nabla^2 \tilde V$ (where $\tilde V$ is the wall-normal velocity component of the disturbance) is expressed in \cite{orr-1907} as 
$S=\phi^{1/3}\left[ A \mathcal{J}_{1/3}(\phi) + B \mathcal{J}_{-1/3}(\phi)\right], \ \phi=(C_1 y + C_2)^{3/2} $, where $\mathcal{J}_n$ indicates the Bessel function and  $A,B$ are to be found through the boundary conditions. The formula for $S$ is readily simplified to an expression containing the Airy function through equation 10.4.15 at page 447 in \cite{abramowitz-stegun-1964}.

%------------------------------
\subsection{Derived quantities}
\label{sec:derived-quantities}

Knowledge of the analytical expression for the GSL allows computing a few quantities that will be used in \S\ref{sec:turbulent} when studying the turbulent flow. One such quantity is the GSL thickness $\delta^*$, defined as the location $y^*$ at which the maximum spanwise velocity reduces to $\rme^{-1} A^*$. In the case of turbulent flow, the maximum phase-averaged spanwise velocity will be used to define $\delta^*$.
Another quantity of interest is the power $\PR$ required to generate the wall waves, expressed as the percentage of the power used to drive the fluid along the streamwise direction in the fixed-wall configuration: 
\begin{equation}
\mathcal{P}_{req}(\%) = \frac{100}{\lambda_x^* T^* \utau U_b^*}
\int_0^{\lambda_x^*} \int_0^{T^*} \pipe{ w_w^* \frac{\p w^*}{\p y^*}}_{y^*=0}
\mbox{d} t^* \mbox{d} x^*,
\label{eq:psp}
\end{equation}
where $T^* =2 \pi/\omega^*$ is the period of the wall forcing. By use of (\ref{eq:airy-solution}), it follows that
\begin{equation}
\label{eq:psp-gsl}
\mathcal{P}_{req,\ell}(\%)= \frac{100 (A^*)^2}{2 \utau U_b^*} \Re \left\{ C \rme^{\pi i/6} 
\left( \frac{\kappa_x^* \utau}{\nu^*} \right)^{1/3} \mbox{Ai}^\prime 
\left[ - \rme^{\pi i/6} \left( \frac{\kappa_x^* \utau}{\nu^*} \right)^{1/3}
\left( \frac{\omega^*}{\kappa_x^* \utau} + \frac{i \kappa_x^* \nu^*}{\utau} \right) 
\right]
\right\},
\end{equation}
where the prime indicates the first derivative of the Airy function and the subscript $\ell$ henceforth indicates a laminar quantity. For $\kappa_x^*=0$, use of (\ref{eq:stokes-flow}) leads to $\mathcal{P}_{req,\ell}(\%)=100 (A^*)^2 (\pi /(\nu^* T^*))^{1/2}/(2 U_b^* \utau)$.

%%%%%%%%%%%%%%%%%%%%%%%%%%%%%%%%%%%%%%%%%%%%%%%%%%%%%%%%%%%%%%%%%%%%%%%%%
\section{Turbulent flow}
\label{sec:turbulent}

The turbulent channel flow forced by the wall waves (\ref{eq:space-time}) is now considered. The results of the laminar analysis will be used throughout this section.

%%%%%%%%%%%%%%%%%%%%%%%%%%%%%%
\subsection{The scaling issue}
\label{sec:turbulent-scaling}

The issue of scaling the wall-forcing parameters and the turbulent statistics is addressed first. In previous studies on large-scale spanwise forcing for drag reduction, different approaches have been followed. Quantities have sometimes been scaled through outer units, i.e. the bulk streamwise velocity and the half-channel height \citep[see for example][]{jung-mangiavacchi-akhavan-1992,quadrio-sibilla-2000}; in other works, viscous units (based on $\nu^*$ and $u_\tau$) have been employed, computing $u_\tau$ either from the reference flow \citep{laadhari-skandaji-morel-1994,xu-huang-2005,itoh-etal-2006} or from the drag-reduced flow \citep{baron-quadrio-1996,choi-2002,ricco-wu-2004}. The different choices for nondimensionalization of flow statistics are perhaps one of the reasons why there is no general consensus on how the oscillating wall modifies the turbulence to reduce drag.

We therefore aim to clarify this point by expanding the DNS database in QRV09, computed at constant mass flow rate $Q$, with a new set of simulations carried out at constant mean streamwise pressure gradient $P_x$. (All the other parameters are kept equal to those in QRV09, to which the reader is referred for details on the numerical procedures.) When $P_x$ is kept constant, the flow rate may change as a consequence of the wall motion, but the friction Reynolds number remains fixed at $Re_\tau=200$: an unequivocal wall-units scaling is defined. For both the constant-$Q$ and the constant-$P_x$ flows, the drag reduction, denoted by $DR_{Q}$ and $DR_P$ respectively, is defined as the percent change in skin-friction coefficient, $C_f= 2 \nu^* \mbox{d} \avg{u^*}/ \mbox{d}y^*|_{y^*=0}/ (U_b^*)^2$. (The symbol $\avg{ \cdot }$ indicates averaging over time and along the homogeneous directions $x$ and $z$.) $DR_{Q}$ is caused by a change in wall-shear stress, whereas $DR_P$ is produced by a change in mass flow rate.

\begin{figure}
\centering
\psfrag{k}{$\kappa_x^+$}
\psfrag{o}{$\omega^+$}
\centering
\includegraphics[width=\textwidth]{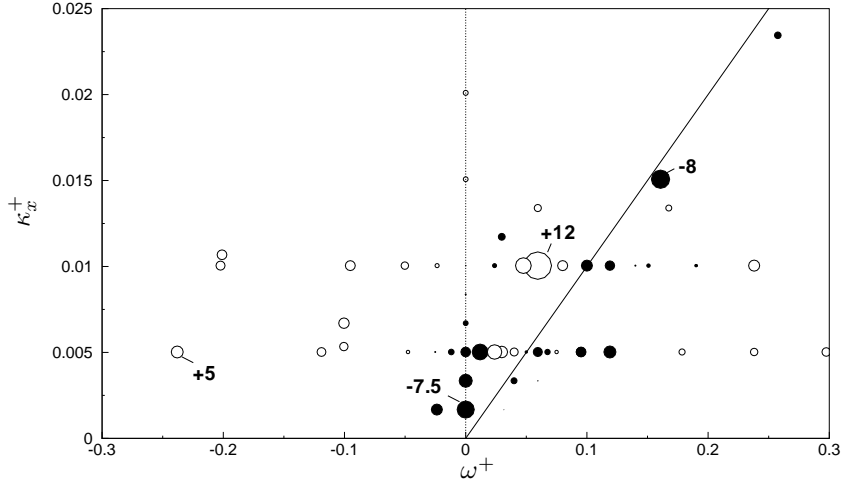}
\caption{Difference between drag reduction $DR_P$ measured for constant pressure gradient $P_x$ and $DR_Q$ for constant flow rate $Q$, at $A^+=12$. White simbols: $DR_P > DR_Q$; black symbols: $DR_P < DR_Q$. Symbol size is proportional to $|DR_P - DR_Q|$. Straight line is $\Ut^+ = 10$.}
\label{fig:DRdiff-Pxconst-Qconst}
\end{figure}

We now study the difference between $DR_P$ and $DR_Q$ to investigate whether the inner-unit scaling holds. Figure \ref{fig:DRdiff-Pxconst-Qconst} shows $DR_P-DR_Q$ as a function of $\kappa_x^+$ and $\omega^+$. The symbol size is proportional to the absolute value of the difference, white symbols denote $DR_P > DR_Q$ and black symbols indicate $DR_P < DR_Q$. Constant-$Q$ data, scaled by $u_\tau$ of the reference flow, are from QRV09; linear interpolation is used when the exact combination of $\kappa_x^+$ and $\omega^+$ is not available. Although the $DR_Q$ and $DR_P$ data follow the same qualitative trend, and the maximum drag reduction is essentially unchanged, quantitative differences are noticeable. A first reason is that the forcing amplitude in the two datasets is actually different, since constant-$Q$ data have larger $A^+$ when drag is reduced and smaller $A^+$ when drag is increased. However, this effect is not particularly intense since the curve of $DR$ vs $A^+$ almost saturates at such high values of $A^+ \approx 12$ (see figure 6a in QRV09). The largest differences are observed to be located near the $\omega^+ / \kappa_x^+ = 10$ line, where the $DR$ gradients with respect to $\kappa_x^+$ and $\omega^+$ are largest. Constant-$P_x$ data consistently indicate larger drag reductions in the high-frequency region (hence white dots). This is because the constant-$Q$ data correspond to larger $\omega^+$, when properly scaled by $u_\tau$ of the modified flow; as a consequence, smaller $DR_Q$ are measured because drag reduction decreases as the frequency increases. Lastly, on the $\kappa_x^+$ axis and near the origin, there is a threshold value of $\kappa_x^+$ below which drag reduction quickly decreases; this value, when expressed in wall units, increases in constant-$Q$ simulations (hence black dots), where a lower $u_\tau$ is measured. These discrepancies support the idea that the drag-reduced flows at constant $P_x$ and constant $Q$ are equivalent when scaled through inner units of the modified flow.  

\begin{figure}
\centering
\psfrag{y}{$y^+$}
\psfrag{u}{$u_{rms}^+, v_{rms}^+, w_{rms}^+$}
\psfrag{o}{$\Omega_{x,rms}^+, \Omega_{y,rms}^+, \Omega_{z,rms}^+$}
\centering
\includegraphics[width=\textwidth]{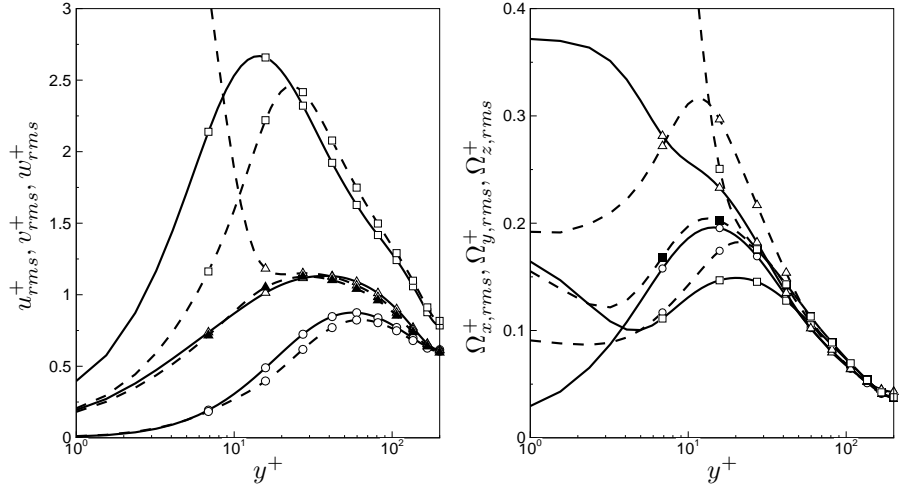}
\caption{Wall-normal distributions of r.m.s. value of velocity (left) and vorticity (right) fluctuations. Symbols: squares, streamwise component; circles, wall-normal component; triangles, spanwise component. Solid lines are for the reference case, and dashed lines are for the traveling-wave case with $A^+=12$, $\omega^+=0.045$, and $\kappa^+=0.012$ ($DR_P$=45\%). The curves with filled symbols are computed by removing the GSL flow.}
\label{fig:rmsvelvort}
\end{figure}

Turbulence statistics for the reference flow and for the traveling-wave flow at constant $P_x$ with $A^+=12$, $\omega^+=0.045$, and $\kappa^+=0.012$ ($DR_P$=45\%) are now studied. Figure \ref{fig:rmsvelvort} shows the wall-normal distributions of the r.m.s. value for the fluctuating velocity and vorticity components. The $u_{rms}^+$ profile reduces up to $y^+\approx20$, and increases slightly for $20 < y^+ < 100$. The most evident change is the upward shift of its peak by about 8 wall units. This behaviour is consistent with previous results for the oscillating wall \citep{baron-quadrio-1996,ricco-wu-2004} where quantities have been scaled by the drag-reduced friction velocity. More substantial reductions of turbulent fluctuations are obviously brought forward by other studies where the friction velocity of the reference flow is used for wall-units scaling (which is equivalent to outer-units scaling); these large changes should be attributed purely to the scaling of choice. The $v_{rms}^+$ profile is unaffected up to $y^+=7$ and mildly attenuated up to $y^+=100$; the $w_{rms}^+$ profile is also largely unvaried, except of course for the thin near-wall region where the GSL oscillations are significant. If the GSL flow is removed (curve with filled triangles), the $w_{rms}^+$ profile becomes nearly identical to that of the reference flow.

On the right of figure \ref{fig:rmsvelvort}, r.m.s. profiles of vorticity fluctuations are plotted. Previous works \citep[see for example][]{karniadakis-choi-2003} have linked changes in the fluctuating vorticity field to the physics of drag reduction by spanwise forcing, focussing primarily on the reduction of streamwise vorticity fluctuations, shown by the squared symbols in figure \ref{fig:rmsvelvort}. Once the GSL is removed (curve with filled squares), $\Omega_{x,rms}^+$ is clearly found to increase up to $y^+\approx70$ when the proper inner scaling is adopted. Even without removing the GSL flow, $\Omega_x^+$ fluctuations increase well beyond the near-wall region where the GSL exerts its influence, which can be estimated to extend to $y^+=15-20$ (see differences between open and closed triangles in figure \ref{fig:rmsvelvort} (both right and left)). This demonstrates that the reported reduction is not directly connected to drag reduction, but probably just an effect of the different Reynolds number. The wall-normal and the spanwise components present opposite trends near the wall. The vorticity $\Omega_{y,rms}^+$ increases with respect to the fixed-wall configuration for $y^+<3$ (direct effect of the forcing) and decreases beyond this location, whereas $\Omega_{z,rms}^+$ is attenuated for $y^+<7$ and presents a maximum at $y^+\approx10$. They are both largely unaffected beyond $y^+\approx50$. 

\begin{figure}
\psfrag{y}{$y^+$}
\psfrag{a}{$a_1$}
\psfrag{U}{$-\avg{uv}^+$}
\psfrag{R}{$R_{uv}$}
\centering
\includegraphics[width=\textwidth]{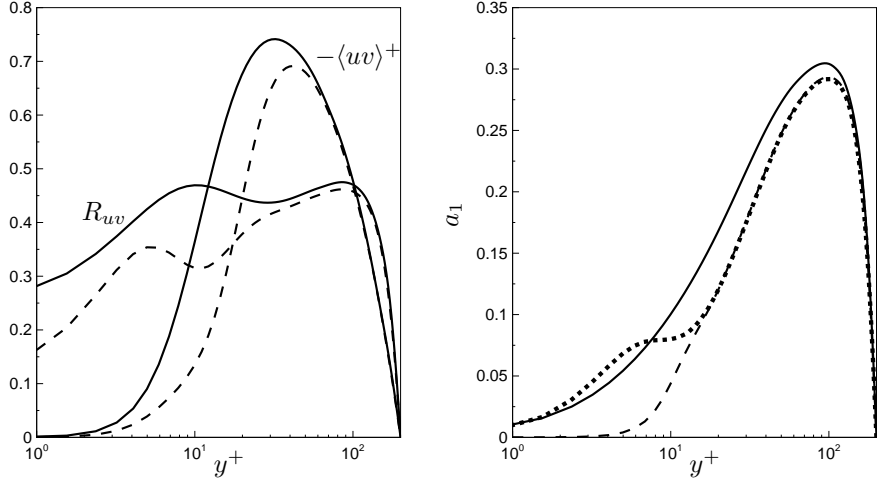}
\caption{Left: wall-normal distributions of Reynolds stress component $-\avg{uv}^+$ and correlation coefficient $R_{uv}$. Right: wall-normal distribution of Reynolds stress structure parameter $a_1$. Flow conditions and lines are as in figure \ref{fig:rmsvelvort}, and the dotted line denotes data computed by removing the GSL flow.}
\label{fig:uvstruct}
\end{figure}

We close this section by presenting quantities that further evidence a structural change of the turbulence throughout the whole channel. The Reynolds stresses $\avg{uv}^+$, plotted in figure \ref{fig:uvstruct} (left), are reduced up to $y^+\approx60$ and the peak moves upward by about 8 wall units. (These profiles cannot be directly related to drag reduction, as the identity by \cite{fukagata-iwamoto-kasagi-2002} only applies to constant-$Q$ flows.) Analogously, the correlation coefficient $R_{uv}=-\avg{uv}^+/(u_{rms}^+ v_{rms}^+)$ presents substantial reductions up to $y^+ \approx 30$, indicating that the Reynolds stresses $\avg{uv}^+$ attenuate more substantially than the single-component r.m.s. values where the viscous effects of the GSL are relevant. 

The Reynolds stress structure parameter 
\[
a_1= \frac{2 \sqrt{\avg{uv}^{+ 2}+\avg{uw}^{+ 2}+\avg{vw}^{+ 2}}}
{\left(u_{rms}^{+ 2} + v_{rms}^{+ 2} + w_{rms}^{+ 2}\right)}
\]
is shown in figure \ref{fig:uvstruct} (right): it is significantly affected by the wall waves. Similarly to the non-equilibrium spanwise-sheared wall-bounded flow studied by \cite{coleman-kim-le-1996}, the Reynolds stresses are attenuated more significantly than the total contribution of the r.m.s. of the velocity fluctuations up to $y^+ \approx 100$. By removing the GSL flow, it emerges that the strong near-wall decrease is simply due to the increase of $w_{rms}^+$ immediately above the waves. Note that when the GSL flow is excluded, the only contribution to $a_1$ is due to $\avg{uv}^+$ because $\avg{uw}^+$ and $\avg{vw}^+$ are null.

Overall, it appears that a proper and consistent scaling of turbulence statistics is required for such flows, where drag reduction may be so high that the friction Reynolds number is significantly changed. 

%%%%%%%%%%%%%%%%%%%%%%%%%%%%%%%%%%%%%%%%%%%%%%%%%%%%%%%%%%%%%%%%%%%%%%%%%%%%%%%%%%%%%
\subsection{Comparison between laminar GSL and turbulent spanwise flow}
\label{sec:turbulent-comparison}
We now ask ourselves whether the GSL solution (\ref{eq:airy-solution}) may be used to describe the space-averaged spanwise turbulent velocity profile. To this aim, we study the spanwise velocity profiles as function of $y^+$ at various phases, the boundary layer thickness $\delta^+$, and the power $\PR$ required to enforce the waves (see \S\ref{sec:derived-quantities}). Comparing the values of $\delta^+$ verifies the agreement in the outer portion of the boundary layer, while $\PR$ is relevant at the wall because it is proportional to the spanwise component of wall friction. 

\begin{figure}
\centering
\psfrag{w}{$w^+$}
\psfrag{y}{$y^+$}
\includegraphics[width=\textwidth]{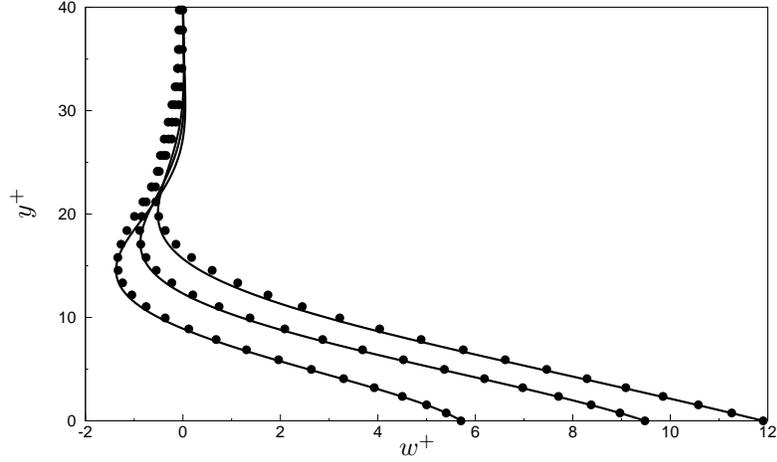}
\caption{Comparison, at three different phases of the cycle, between the GSL velocity profile computed via (\ref{eq:airy-solution}) (lines), and the mean spanwise turbulent profile computed by QRV09 (symbols), for $A^+=12$, $\kappa_x^+=0.0084$ and $\omega^+=0.03$.}
\label{fig:comparison}
\end{figure} 
Figure \ref{fig:comparison} shows the comparison between the laminar profiles computed by (\ref{eq:airy-solution}) with the turbulent space-averaged profiles for $A^+=12$, $\kappa_x^+=0.0084, \omega^+=0.03$ ($DR_P \approx 48\%$). The agreement is very good, except for small discrepancies at the outer edge of the layer. 

\begin{figure}
\centering
\psfrag{t}{$\delta_t^+$}
\psfrag{l}{$\delta_\ell^+$}
\includegraphics[width=\columnwidth]{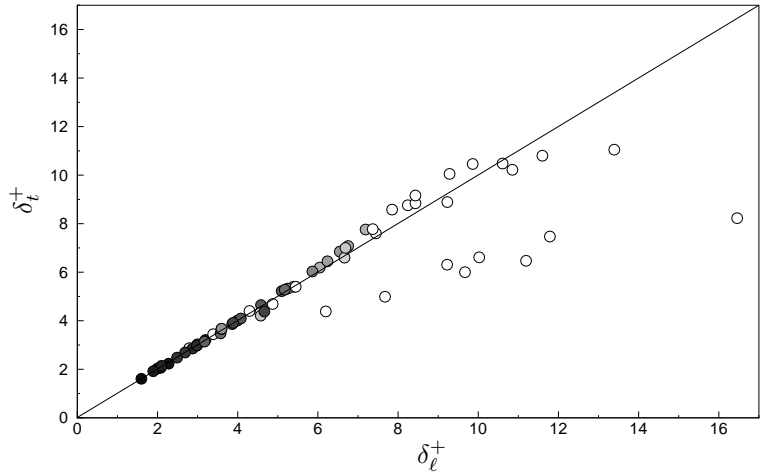}
\caption{Comparison between $\delta^+_\ell$ computed by (\ref{eq:airy-solution}) and $\delta^+_t$ from constant-$P_x$ DNS. Here and in the following figures, the grey scale indicates $\T^+$: the colour changes on a linear scale from black at $\T^+=0$ to white at $\T^+ \ge T_{th}^+=120$. White points off the linear correlation are those for which $\T^+$ becomes so large that the flow experiences drag increase.} 
\label{fig:error-delta}
\end{figure}
The values of the GSL thicknesses $\delta_\ell^+$ and $\delta_t^+$, for the laminar and turbulent flows respectively, are plotted in figure \ref{fig:error-delta}. The thickness $\delta_\ell^+$ is extracted from the analytical solution (\ref{eq:airy-solution}), whereas $\delta_t^+$ is obtained from the constant-$P_x$ DNS. (The dataset produced by QRV09 comprised a larger number of constant-$Q$ simulations, but did not contain the information required to compute $\delta_t^+$.) Most of the points show excellent agreement, although a few of them are far from the line $\delta_t^+ = \delta_\ell^+$. Although the reason for this behaviour was already hinted at in QRV09, it is worth discussing it further here, since this concept will be useful to explain several results in the following. We resort to the concept of period of oscillation $\T^+ \equiv |\lambda_x^+/(\Ut^+-\Uw^+)|$, introduced by QRV09 to study the physics of the traveling waves. $\T^+$ is the period of oscillation as seen by an observer moving at the same speed $\Uw^+$ of turbulence fluctuations. When $\T^+ \gg T_{th}^+ = 120$ (where the threshold value $T_{th}^+$ is linked to the life time of the near-wall turbulent coherent structures, i.e. an auto-correlation time observed in a Lagrangian frame, see also \cite{quadrio-ricco-2004}), the spanwise forcing becomes too slow and couples directly with the streamwise flow. In the following, the terminologies small and large $\T^+$ are to be intended with respect to $T_{th}^+$. The stramwise flow becomes highly distorted and ultimately experiences an increase of drag (see figure 7 of QRV09). In figure \ref{fig:error-delta}, the data points are coloured in grey scale, with darker hues corresponding to smaller $\T^+$ and white points indicating $\T^+ > 120$. Excellent agreement between $\delta_\ell^+$ and $\delta_t^+$ is found for $\T^+ < T_{th}^+$. All the white large-$\T^+$ circles lying below the straight line correspond to drag-increase cases. Therefore, $\delta_\ell^+$ and $\delta_t^+$ show very good agreement as long as the wall forcing is unsteady enough to avoid strong coupling with the streamwise flow.

\begin{figure}
\centering
\psfrag{t}{$\mathcal{P}_{req,t}(\%)$}
\psfrag{l}{$\mathcal{P}_{req,\ell}(\%)$}
\includegraphics[width=\columnwidth]{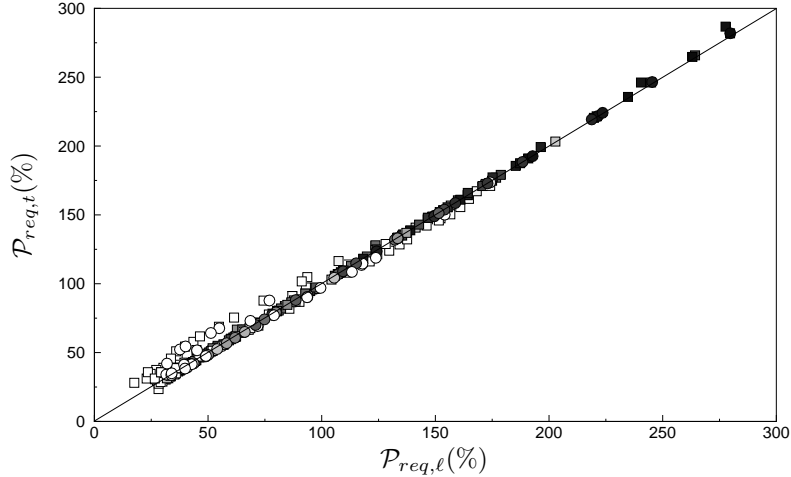}
\caption{Comparison between the required power $\mathcal{P}_{req,\ell}(\%)$ computed by (\ref{eq:psp-gsl}) and $\mathcal{P}_{req,t}(\%)$ obtained from DNS at $A^+=12$. Squares indicate constant-$Q$ cases from QRV09, while circles indicate constant-$P_x$ calculations. The grey scale is as in figure \ref{fig:error-delta}.}
\label{fig:error-psp}
\end{figure}
The same conclusion is arrived at by looking at the power $\PR$ required to enforce the waves (figure \ref{fig:error-psp}). The power $\mathcal{P}_{req,\ell}$, given by (\ref{eq:psp-gsl}), is compared with the DNS turbulent $\mathcal{P}_{req,t}$, computed through (\ref{eq:psp}) where the mean velocity gradient at the wall $\mbox{d} \avg{u^*} /\mbox{d}y^*|_{y^*=0}$ is used instead of $\utau$. Again, the dark points corresponding to small $\T^+$ fall on the line $\mathcal{P}_{req,\ell} = \mathcal{P}_{req,t}$, whereas points far from the line are mostly white. Note that in figure \ref{fig:error-psp} the QRV09 dataset is displayed, too. Here and in the following figures, the forcing parameters at constant $Q$ are rescaled through inner units of the modified flow. As there is no evident difference between the two datasets, it can be concluded that the above observations hold regardless of the flow being at constant flow rate or driven by a constant pressure gradient.

The good agreement between the laminar and the turbulent profiles can be better understood by studying the turbulent spanwise momentum equation. By decomposing the velocity field as:
\[
{\bf u}^+ (x^+,y^+,z^+,t^+) = 
\left\{\overline{U}^+(\xi^+,y^+),0,\overline{W}^+(\xi^+,y^+) \right\} 
+ \{u',v',w'\},
\]
where the overbar indicates quantities averaged in time and along the $z$ direction, the spanwise momentum equation reads:
\[
\left( \overline{U}^+ - U_t^+ \right) \frac{\p \overline{W}^+}{\p \xi^+} 
- \frac{\p^2 \overline{W}^+}{\p \xi^{+ 2}} - \frac{\p^2 \overline{W}^+}{\p y^{+ 2}} =
- \frac{\p \overline{U}^+}{\p \xi^+} \overline{W}^+
- \frac{\p (\overline{u'w'}^+)}{\p \xi^+}  
- \frac{\p (\overline{v'w'}^+)}{\p y^+}.
\]

The terms on the l.h.s. coincide with the ones in the laminar equation, while the terms on the r.h.s. become non-zero only in the turbulent case, and their magnitude gives an indication of the difference between the solutions of the laminar and turbulent equations. Computing the various terms, with time and $z$-averaging, for a typical dark point in figure \ref{fig:error-delta} indicates that the largest l.h.s. term, namely the viscous term $\p^2 \overline{W}^+ / \p y^{+ 2}$, exceeds the largest r.h.s. term, i.e. the Reynolds stress term $\p (\overline{v'w'}^+) / \p y^+$, by more than two orders of magnitude.
Therefore, $\overline{W}$ closely obeys the laminar equation in the turbulent regime, too. This confirms the explanation already given by \cite{ricco-quadrio-2008} to the agreement between laminar and turbulent spanwise profiles for the oscillating-wall case. 

%%%%%%%%%%%%%%%%%%%%%%%%%%%%%%%%%%%%%%%%%%%%%%%%%%%%%%%%%%%%%%%%%%%
\subsection{Laminar quantities and turbulent drag}
\label{sec:turbulent-correlation}

The laminar solution is now used to relate the changes of turbulent drag to quantities computed from the analytical solution (\ref{eq:airy-solution}). 

\subsubsection{Role of $\delta^+$}
The thickness of the spanwise layer has been recognized as an important quantity for the oscillating wall flow \citep[see for example][]{choi-xu-sung-2002}. The occurrence of an optimal thickness for drag reduction has been explained through the effectiveness of the viscous shearing action of the moving wall to weaken the interactions between the near-wall streaks and the vortical structures \citep{karniadakis-choi-2003}. QRV09 observed that, for maximum drag reduction induced by the oscillating wall, the standing waves, and the traveling waves, the GSL thickness showed very similar values, i.e. $\delta^+ \approx 6.5$. As discussed in \S\ref{sec:turbulent-scaling}, the changes of turbulent statistics near the wall, i.e. where the GSL viscous effects are relevant, further suggest that the spanwise viscous layer may be linked to the structural change of the turbulent flow and therefore to drag reduction. The above observations prompt us to study the dependence of $\delta^+$ on the forcing parameters and to explore the relation of $\delta^+$ with drag reduction.

\begin{figure}
\centering
\psfrag{k}{$\kappa_x^+$}
\psfrag{o}{$\omega^+$}
\centering
\includegraphics[width=\textwidth]{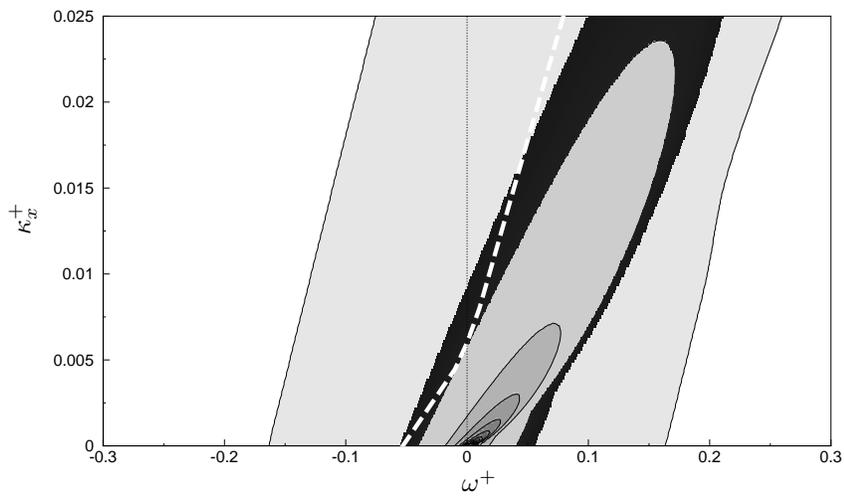}
\caption{Laminar GSL thickness $\delta^+_\ell$ as function of $\kappa_x^+$ and $\omega^+$, computed by (\ref{eq:airy-solution}). Contour levels start from $\delta^+_\ell=3.5$ with an increment of 3.5. The white dashed line represents the locus of maximum $DR$ at fixed $\kappa_x^+$ as extracted from QRV09. See text for further discussion on this line and the black region defined by $6 < \delta^+_\ell < 7$.}
\label{fig:k-omega-deltalam}
\end{figure}

The laminar GSL thickness $\delta^+_\ell=\delta^+_\ell(\kappa_x^+,\omega^+)$ is computed through (\ref{eq:airy-solution}) (when $\kappa_x^+=0$, the Stokes layer thickness is computed by $\delta^+_\ell= \sqrt{2 / \omega^+}$) and shown in figure \ref{fig:k-omega-deltalam}. Large values of $\delta^+_\ell$ are found in the first quadrant, and therefore pertain to forward-traveling waves. The thickness is large near the origin, but quickly drops as either $\omega^+$ or $\kappa_x^+$ increases. The map bears an evident similarity with the drag reduction contour plot of figure 2 in QRV09. At large enough frequencies, contour lines tend to become oblique and to align vertically, similarly to the constant-$DR$ lines. The region where $6 < \delta^+_\ell < 7$, indicated by the black area in figure \ref{fig:k-omega-deltalam}, matches well the locus of points with largest $DR$ at fixed $\kappa_x^+$. This locus is taken from the QRV09 data, and is graphically represented by the white dashed line. The match holds for $\omega^+ < 0.05$ (or $\kappa_x^+ < 0.015$). For higher $\omega^+$ (or higher $\kappa_x^+$), the maximum-$DR$ line encounters values of $\delta_\ell^+$ smaller than 6 because the drag-increase region interferes with the black region there (see later \S\ref{sec:turbulent-regimes}). 

\begin{figure}
\centering
\psfrag{D}{$DR(\%)$}
\psfrag{d}{$\delta^+_\ell$}
\psfrag{m}{$\delta^+_{min}$}
\psfrag{o}{$\delta^+_{opt}$}
\centering
\includegraphics[width=\columnwidth]{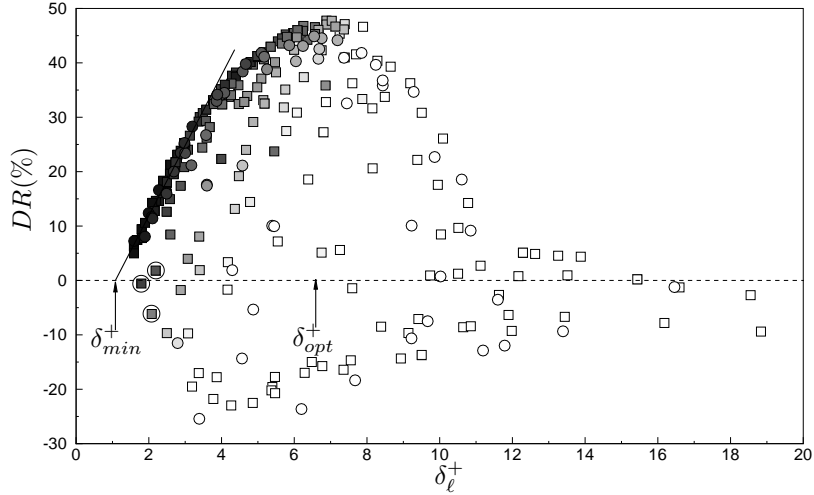}
\caption{Drag reduction data as function of $\delta^+_\ell$ for constant-$Q$ (squares) and constant-$P_x$ (circles) simulations. The oblique straight line shows the linear correlation between $DR$ and $\delta^+_\ell$. The arrows indicate the minimal condition for drag reduction, $\delta^+_\ell \approx 1 = \delta^+_{min}$, and the optimal GSL thickness, $\delta^+_\ell \approx 6.5 = \delta^+_{opt}$. Grey scale is as in figure \ref{fig:error-delta}. See text for discussion on circled points.} 
\label{fig:DR-deltalam}
\end{figure}

In figure \ref{fig:DR-deltalam}, the $DR_P$ data (circles) described in \S\ref{sec:turbulent-scaling} and the $DR_Q$ data by QRV09 (squares) are plotted as function of $\delta^+_\ell$. As in figure \ref{fig:error-delta}, black and grey points correspond to small $\T^+$, i.e. to a wall forcing which is unsteady with respect to the near-wall turbulence, $\T^+ < T_{th}^+$, and white points are for $\T^+>T^+_{th}$. White points are extremely scattered, while black points collapse on a sharply-defined curve. Intermediate grey points with $\T^+ \approx T^+_{th}$ are confined between the small- and large-$\T^+$ ones. At small $\T^+$, when the GSL profile matches the mean spanwise turbulent profile, $DR$ grows linearly with $\delta^+_\ell$. Linearity holds up to $DR(\%) \approx 35$ and $\delta^+_\ell \approx 4$, which confirms the visual analogy at small $\T^+$ between the drag reduction map by QRV09 and the  $\delta^+_\ell$ map in figure \ref{fig:k-omega-deltalam}. The maximum drag reduction occurs for $\delta^+_\ell \approx 6.5 = \delta^+_{opt}$ (see arrow in figure \ref{fig:DR-deltalam}), as already emerged when studying figure \ref{fig:k-omega-deltalam}. Note that, as expected, the grey points at maximum drag reduction correspond to $\T^+ \approx T^+_{th}$, i.e. they exist on the border of the oblique strip $\T^+ \leq T^+_{th}$ described in QRV09. 

\begin{figure}
\centering
\psfrag{D}{$DR(\%)$}
\psfrag{d}{$\sqrt{\delta^+_\ell}$}
\centering
\includegraphics[width=\columnwidth]{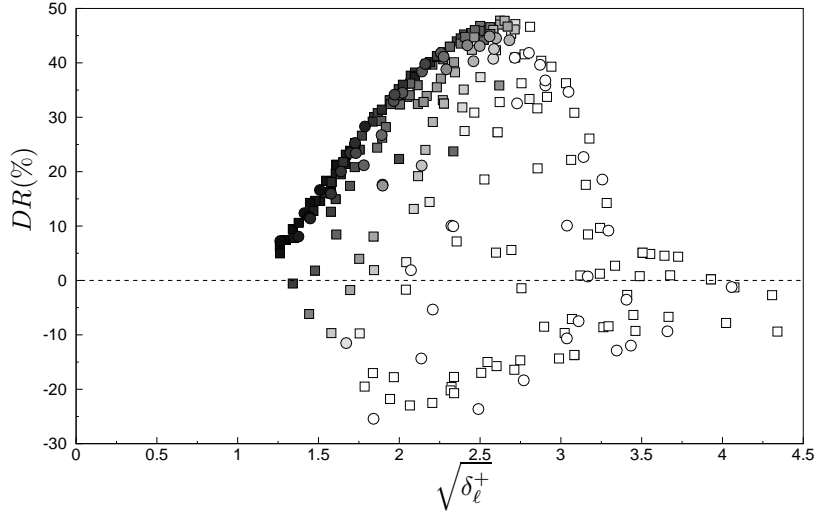}
\caption{Drag reduction data as function of $\sqrt{\delta^+_\ell}$ for constant-$Q$ (squares) and constant-$P_x$ (circles) simulations. Grey scale is as in figure \ref{fig:error-delta}.} 
\label{fig:DR-sqrt-deltalam}
\end{figure}

The dataset of figure \ref{fig:DR-deltalam} is replotted in figure \ref{fig:DR-sqrt-deltalam} by using the quantity $\sqrt{\delta^+_\ell}$ on the horizontal axis. A similar, perhaps improved, collapse of the data on a straight line is observed, although it is difficult to discriminate between $DR \sim \sqrt{\delta^+_\ell}$ and $DR \sim \delta_\ell^+$ up to $\delta_\ell^+ \approx 4$ and saturating at higher values. 

It must be observed that, in both cases, a few dark points with small $\delta^+_\ell$ (circled points in figure \ref{fig:DR-deltalam}) do not correlate well: these points correspond to high values of $\kappa_x^+$, where the drag-increase region extends outside the strip $\T^+ \leq T^+_{th}$. We thus conclude that $DR$ is related to $\delta^+_\ell$ as long as $\T^+ \ll T^+_{th}$ (dark points) and $|\Ut^+ - \Uw^+| \gg 0$. (For the oscillating-wall regime, $\T^+= |T^+| \gg T^+_{th}$). Under these conditions, the laminar solution can therefore be used effectively to predict $DR$. The reader is referred to \S\ref{sec:turbulent-regimes} for further discussion on the classification of the flow regimes induced by the wall forcing.

%----------------------------------------------------
\subsubsection{Minimal conditions for drag reduction}

A further observation can be made from figure \ref{fig:DR-deltalam}. When extrapolated to small values of $\delta_\ell^+$, the curve where the dark points collapse crosses the zero-$DR$ line at a non-zero abscissa, $\delta_\ell^+ \approx 1$, suggesting that a finite thickness is needed to yield drag reduction. We refer to this thickness as the {\em minimal} GSL thickness $\delta^+_{min}$ for drag reduction (indicated by an arrow in figure \ref{fig:DR-deltalam}). This concept has already been advanced by \cite{ricco-quadrio-2008} for the oscillating-wall technique, and it can be extended to finite values of drag reduction. A unique minimal value of $\delta^+_\ell$ (for which $\T^+ \ll T^+_{th}$) must be enforced to obtain a specific amount of drag reduction. For example, figure \ref{fig:DR-deltalam} shows that $\delta^+_\ell > 2.5$ is needed to obtain $DR > 20\%$, and $\delta^+_\ell > 4.5$ is necessary for $DR>40\%$. The occurrence of the minimal conditions is even more evident when the $DR$ dataset is plotted as function of $\sqrt{\delta^+_\ell}$ in figure \ref{fig:DR-sqrt-deltalam}.

%--------------------------------
\subsubsection{Role of $\PR$}

\begin{figure}
\centering
\psfrag{k}{$\kappa_x^+$}
\psfrag{o}{$\omega^+$}
\centering
\includegraphics[width=\textwidth]{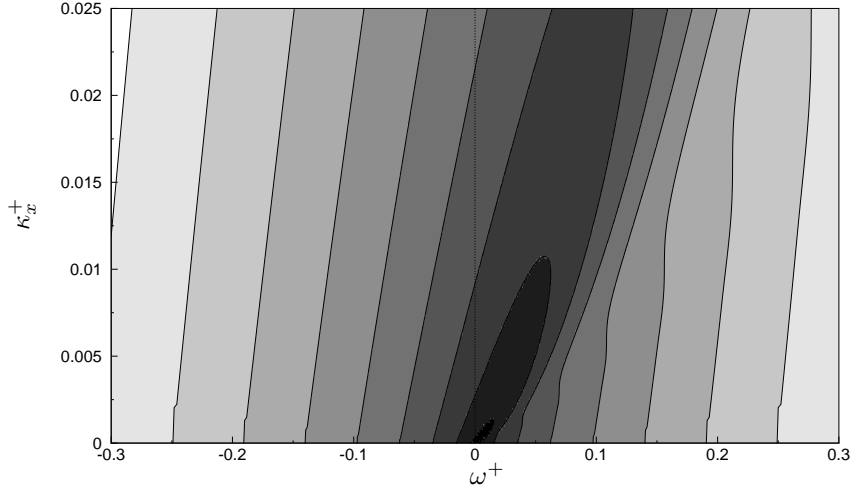}
\caption{Power $\mathcal{P}_{req,\ell} (\%)$ required to impose the traveling waves, computed by the analytical expression (\ref{eq:psp-gsl}). Contours start at 20\% near the origin and increase by 20\% steps.}
\label{fig:k-omega-psplam}
\end{figure}

Figure \ref{fig:k-omega-psplam} shows $\mathcal{P}_{req,\ell}=\mathcal{P}_{req,\ell}(\kappa_x^+,\omega^+)$, computed by (\ref{eq:psp-gsl}). The contour plot is qualitatively similar to the $\delta_\ell^+$ map in figure \ref{fig:k-omega-deltalam}; this is expected, since both quantities represent the viscous diffusion from the wall. 
\begin{figure}
\centering
\psfrag{D}{$DR(\%)$}
\psfrag{r}{$\mathcal{P}_{req,\ell}$(\%)}
\centering
\includegraphics[width=\columnwidth]{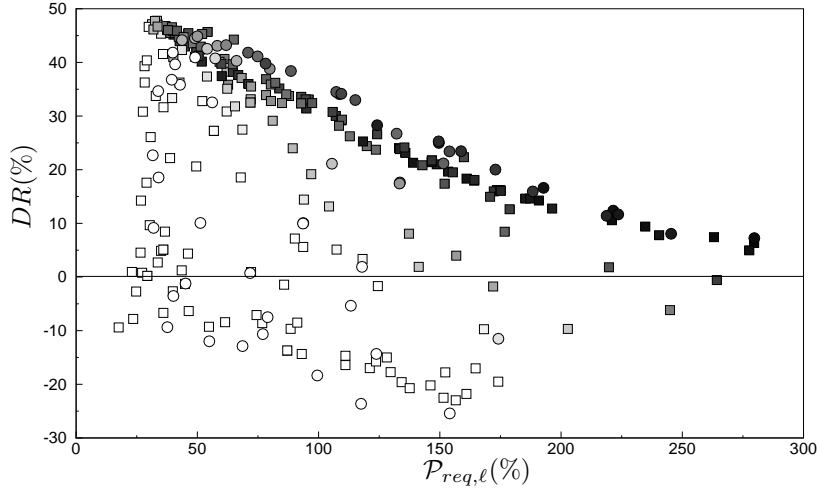}
\caption{Drag reduction data as function of $\mathcal{P}_{req,\ell}$(\%) for constant-$Q$ (squares) and constant-$P_x$ (circles) simulations. Grey scale is as in figure \ref{fig:error-delta}.}
\label{fig:DR-psplam}
\end{figure}

In figure \ref{fig:DR-psplam}, the $DR$ data are plotted versus $\mathcal{P}_{req,\ell}$. Similarly to the analysis with $\delta^+_\ell$, the correlation is good for dark symbols and worsens as $\T^+$ increases. For the dark points, $DR$ decreases monotonically as $\PR$ increases, which is therefore minimum when $DR$ is maximum. This supports quantitatively the observation by QRV09 that the net energy saving produced by the traveling waves can be high. When compared with figure \ref{fig:DR-deltalam}, the collapse is less accurate, with the dark constant-$P_x$ circles showing slightly higher $DR$ values than the black constant-$Q$ squares. This effect is due to the scaling of the forcing amplitude, already mentioned in \S\ref{sec:turbulent-scaling}. For constant-$Q$ simulations, $A^+$ increases when scaled by the friction velocity of the drag-reduced flow, so that $\mathcal{P}_{req,\ell}$, which depends quadratically on $A^+$ (see (\ref{eq:psp-gsl})) is smaller for constant-$Q$ data. This effect is absent in the correlation with $\delta^+_\ell$ in figure \ref{fig:DR-deltalam} because $\delta^+_\ell$ is independent of $A^+$ by definition.

%%%%%%%%%%%%%%%%%%%%%%%%%%%%%%%%%%%%%%%%
\subsection{Four regimes for drag modification}
\label{sec:turbulent-regimes}
\begin{figure}
\psfrag{U}{$\Ut^+=\Uw^+$}
\psfrag{TeqTth}{$\T^+=T_{th}^+$}
\psfrag{DR0}{$DR(\%)=0$}
\psfrag{1}{\mycirc{\tiny 1}}
\psfrag{2}{\mycirc{\tiny 2}}
\psfrag{3}{\mycirc{\tiny 3}}
\psfrag{4}{\mycirc{\tiny 4}}
\psfrag{kx}{$\kappa_x^+$}
\psfrag{omegae}{$\varpi^+$}

\psfrag{Utabs-Uw}{$\Ut^+ - \Uw^+$}
\psfrag{Ut}{$\Ut^+$}
\psfrag{Uw}{$\Uw^+$}
\psfrag{app}{$\approx$}
\psfrag{0}{$0$}
\psfrag{m0}{$\gg$0}
\psfrag{Teq}{$\T^+ - T_{th}^+$}
\psfrag{lTth}{$\ll 0$}
\psfrag{mTth}{$\gg 0$}
\psfrag{ldi}{Weak drag increase}
\psfrag{rdi}{Lock-in drag increase}
\psfrag{adr}{Active drag reduction}
\psfrag{sdr}{Weak drag reduction}
\centering
\includegraphics[width=0.8\columnwidth]{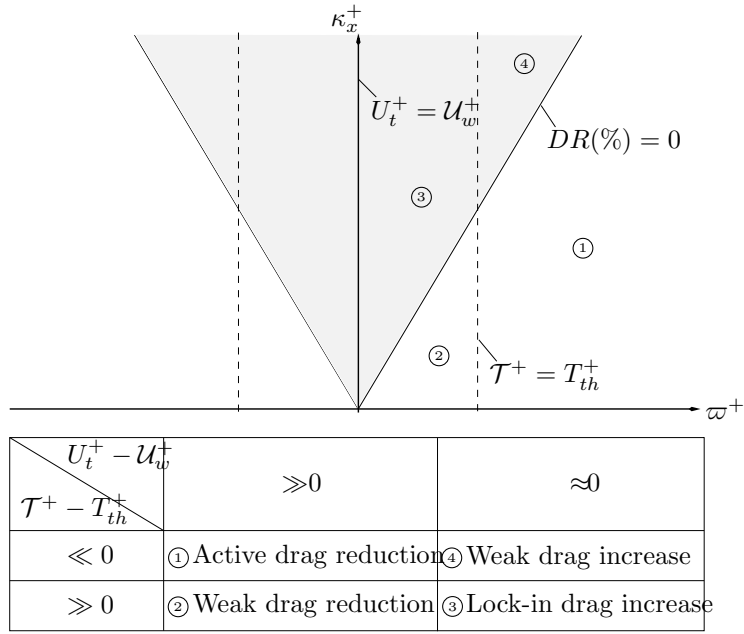}
\caption{Schematic of the four mechanisms by which the traveling waves affect the turbulent friction drag, illustrated as different regions in the $\kappa_x^+ - \varpi^+$ plane. In the shaded area, friction drag is increased. The table at the bottom describes the four numbered regions in the right half-plane, where waves travel faster than near-wall turbulence.}
\label{fig:fourregimes}
\end{figure}

In \S\ref{sec:turbulent-correlation}, it has emerged that the period $\T^+$ is one of the key parameters. It is an index of the unsteadiness of the forcing in a frame moving at the (average) speed of the near-wall turbulence fluctuations. The other important parameter is the wave speed $\Ut^+$. We therefore interpret the drag reduction data in a new coordinate setting, namely $\kappa_x^+$ and $\varpi^+=2 \pi/\T^+$. The quantity $\varpi^+ / \kappa_x^+$ expresses the wave phase speed as seen by an observer traveling at $\Uw^+$ while the quantity $\varpi^+$ is an index of the unsteadiness of the forcing in the convecting reference frame. 

The top graph in figure \ref{fig:fourregimes} shows a schematic of such a map. The strip $\T^+ = T_{th}^+$, inside which the forcing is quasi-steady with respect to the near-wall turbulence, and the cone $\omega^+ / \kappa_x^+ = \Uw^+ \pm 2$, inside which the waves lock with the convecting turbulence and produce drag increase, are now both centered about the $\varpi^+=0$ axis. The first quadrant represents waves travelling forward faster than the near-wall turbulence, i.e. $\Ut^+ > \Uw^+$, while in the second quadrant the waves move either forward or backward and are slower than the turbulence. 

The intersection of the strip (identified by $\T^+$ alone) with the cone (identified by $\T^+$ and $\kappa_x^+$) defines four regions in the first quadrant and four in the second one. We first focus on the first quadrant. In regions 1 and 2, i.e. outside the drag-increase half-cone, the waves move significantly faster than the turbulence. In regions 3 and 4, i.e. inside the half-cone, the waves and the turbulence travel at approximately the same speed. When observed in a frame of reference moving with the waves, a turbulent structure traveling at $\Uw^+$ covers a length longer than one wavelength in regions 1 and 4, and shorter than one wavelength in regions 2 and 3. The four regimes are also schematically described in the table at the bottom of figure \ref{fig:fourregimes}. 

Region 1, termed region of active drag reduction, is outside the strip and outside the half-cone; the GSL thickness $\delta^+$ determines drag reduction. The success of GSL in reducing drag is due to the spanwise viscous forces operating on a shorter time scale than the typical Lagrangian correlation time of near-wall structures, and through wall waves that travel faster than the turbulence structures. The laminar GSL thickness agrees well with the corresponding turbulent thickness, which makes the laminar analysis useful for predicting drag reduction.

Region 2, of weak drag reduction, is inside the strip and outside the half-cone; a sharp drop of drag reduction occurs as $\T^+$ increases beyond the optimum time scale $T_{th}^+$. The forcing is slow with respect to the turbulence, although the waves still travel faster than the structures. The GSL becomes thick and the near-wall turbulence is not efficiently altered because a typical structure loses its coherence before traveling a distance of one wavelength. 

Region 3 is inside both the strip and the half-cone, and corresponds to high drag increase. Region 4 of weak drag increase sits outside the strip and inside the half-cone; not many simulations are available for wave parameters falling into this region. The wave speed is comparable with the one of the near-wall structures, which cover only a small portion of one wavelength during their survival time. The resulting flow field is highly distorted and three-dimensional, as visualized by QRV09 in their figure 7 (bottom). This occurs irrespectively of $\delta^+$, which is consistent with the previous observation on figure \ref{fig:DR-deltalam} that the drag increase is not related to $\delta^+$.

In the second quadrant, four analogous regions can be distinguished. The above qualitative discussion on the relative interaction between the turbulence and the waves still holds, and the amounts of drag change for regimes 2,3,4 are very similar, but higher drag reductions are observed for regime 1 in the second quadrant.
There is no such a variety of regimes in the analogous laminar case because the sole discriminatory factor is the characteristic speed $U_{t,c}^*$. Different cases are distinguished when this speed is compared with the representative velocity of the streamwise laminar flow within GSL, $\Ud^*$, as discussed in \S\ref{sec:laminar}, but there is no time scale comparison because unsteady fluctuations are absent in the laminar case.

\section{Summary}
\label{sec:summary}

This paper has studied how a plane channel flow is modified by streamwise-traveling waves of spanwise velocity applied at the wall. Both the laminar and turbulent streamwise flows have been considered. 

In the laminar case, the wall forcing induces a thin, unsteady and streamwise-modulated transversal boundary layer, the generalized Stokes layer (GSL), which does not affect the streamwise flow. A linear streamwise velocity profile is assumed, which is exact for a laminar Couette flow and a very good approximation for the near-wall region of the laminar Poiseuille flow of interest here. The GSL velocity profile has been expressed in terms of the Airy function of the first kind, thus generalizing the well-known Stokes analytical solution that describes a still fluid over a plane wall in harmonic motion. 

Through asymptotic analysis, a characteristic phase speed of the traveling waves has been found; it is related to the streamwise velocity of the flow at the edge of the GSL, and discriminates amongst three flow regimes: the oscillating-wall, the standing-wave, and the traveling-wave regimes. In the oscillating-wall regime, the phase speed of the waves is much larger than the characteristic speed, and the GSL behaves as the classical Stokes layer. In the opposite standing-wave regime, the phase speed is much smaller than the characteristic speed, and the GSL becomes the steady, spatially-modulated Stokes layer studied by \cite{viotti-quadrio-luchini-2009}. In the intermediate traveling-wave regime, the phase speed is comparable with the characteristic speed. The inertial effects are given by both its unsteadiness and its streamwise modulation.

The turbulent case is fundamentally different as the wall forcing does affect the streamwise flow, inducing either drag reduction or drag increase. The boundary-layer thickness of the space-averaged spanwise turbulent profile agrees well with the GSL laminar thickness when i) the phase speed of the waves is sufficiently different from the near-wall turbulent convection velocity, and ii) when the waves act on a time scale which is significantly smaller than the survival time of the turbulent structures. When the waves move at a speed comparable with the convection velocity, a lock-in effect renders the instantaneous turbulent flow highly three-dimensional, the friction drag increases and the spanwise laminar solution fails to represents the spanwise turbulent flow. When the waves oscillate on a time scale which is larger than the typical lifetime of the near-wall turbulence, the drag reduction decreases substantially, and again the laminar solution loses its validity. The comparison of velocity and time scales has allowed us to identify four distinct turbulent regimes.

The amount of turbulent drag reduction has been shown to correlate with the GSL thickness; the collapse of the data is very good up to $DR \approx$ 35\%. The validity of such correlation is subject to the same two conditions mentioned above, which renders the laminar profile instrumental for describing the turbulent drag reduction, at least for the Reynolds number conditions studied.

%%%%%%%%%%%%%%%%%%%%%%%%%%%%%%%%%%%%%%%%%%%%%%%%%%%%%%%%%%%%%%%%%%%%%%%%%%%
\section*{Acknowledgments}
We acknowledge the interesting discussions with Dr Fulvio Martinelli. We are also thankful to Dr Andrew Walton for his comments on an early draft of this paper.

\bibliographystyle{plain}

\end{document}